\documentstyle[twocolumn,prl,aps]{revtex}

\newcommand{\txt}{\textstyle}

\newcommand{\beq}{\begin{equation}}
\newcommand{\eeq}{\end{equation}}
\newcommand{\ba}{\begin{array}}
\newcommand{\bea}{\begin{eqnarray}}
\newcommand{\ea}{\end{array}}
\newcommand{\eea}{\end{eqnarray}}

\newcommand\comment[1]{ \hbox{[{\it Comment suppressed here.}\/]} }
\newcommand\hide[1]{}

\newcommand{\skipover}[1]{}

\newcommand{\third}{{\txt {1\over 3}}}

\newcommand{\twothirds}{{\txt {2\over 3}}}

\begin{document}                                                
\draft
\preprint{MIT-CTP-xxxx}
\title{Enforced Electrical Neutrality of the Color-Flavor Locked Phase}
\author{Krishna Rajagopal and Frank Wilczek}
\bigskip
\address{
Center for Theoretical Physics\\
Massachusetts Institute of Technology, Cambridge, MA 02139, USA
}
\date{February 9, 2001}
\maketitle
\begin{abstract} 
We demonstrate that quark matter in the color-flavor locked phase
of QCD is rigorously
electrically neutral, despite the 
unequal quark masses, and even in the presence of 
an electron chemical potential.  As long as the strange quark
mass and the electron chemical potential 
do not preclude the color-flavor locked phase,
quark matter is automatically neutral.  No electrons are
required and none are admitted.  
\end{abstract}

If cold, dense quark matter is approximated as noninteracting
up, down and strange quarks, it has long been understood
that equilibrium with
respect to the weak interactions 
together with the relatively large mass of the strange
quark imply that the strange quarks are less
abundant than the other quarks.  Thus, noninteracting quark 
matter is electrically positive and a nonzero density of electrons
is required in order to obtain electrically neutral bulk matter.

Explicitly, weak equilibrium imposes
\begin{equation}\label{weakequilibrium}
\mu_u=\bar\mu-\twothirds \mu_e\,;\ \mu_d=\mu_s=\bar\mu+\third \mu_e
\end{equation}
where $3\bar\mu$ is the chemical potential for baryon number and
$\mu_e$ is that for electron number. 
In the absence
of interactions, the corresponding number densities are 
\begin{equation}\label{naiveN}
N_{u,d}=\frac{1}{\pi^2}\mu_{u,d}^3;\, N_{s}=
\frac{1}{\pi^2}(\mu_s^2-m_s^2)^{3/2};\, N_e=\frac{1}{3\pi^2}(\mu_e)^3.
\end{equation}
(Throughout, we set $m_u=m_d=m_e=0$ and $m_c=m_b=m_t=\infty$.)
Electric neutrality requires
\begin{equation}
\twothirds N_u -\third N_d -\third N_s - N_e=0\ .
\end{equation}
With $m_s=0$, this can be satisfied with $N_u=N_d=N_s$ and $N_e=0$.
Because $m_s>0$, however, noninteracting quark matter 
must have 
$\mu_e>0$ and an electron density $N_e>0$ if it is to
be electrically neutral.
The condition (\ref{weakequilibrium}) cannot be modified
by the strong 
interactions among the quarks. It has long been known, however, that
interactions can modify the 
relations (\ref{naiveN})\cite{FreedmanMcLerran}.
We argue here that interactions modify these
relations {\it qualitatively}: they favor a state
of quark matter in which 
$N_u=N_d=N_s$ and $N_e=0$ even when $m_s\neq 0$ and $\mu_e\neq 0$,
as long as neither is too large.

It is becoming widely accepted 
that at asymptotic densities the ground
state of 
QCD with $m_s=0$ is
the color-flavor locked
(CFL) phase\cite{CFL,OtherCFL,ioffe}.   
In this phase, color gauge symmetry is completely broken, as
are both chiral
symmetry and baryon number (i.e., the material is superfluid).   
The effective coupling is weak (because
QCD is asymptotically
free), and the ground state and low-energy
properties can be determined by adapting methods used in the theory of
superconductivity (BCS theory)\cite{CFL,OtherCFL,ioffe,CFLmesons}.
The CFL phase persists for finite masses, and
even for unequal  masses, so long as the differences are not too 
large\cite{ABR2+1,SW2}.   
It is very
likely the ground state for real
QCD, assumed to be in equilibrium with respect to the weak interaction, over
a substantial range of
densities.

Since the CFL phase occurs at weak
coupling, it might seem
natural to think that, like noninteracting quark
matter,  CFL quark matter with $m_s\neq 0$ is 
electrically positive, and that its
existence in bulk requires a neutralizing electron fluid.   Indeed, this
has been tacitly assumed in
the literature.   The presence of an electron fluid drastically affects the
low-energy dynamics of dense
matter in the CFL phase.   Specifically, for example, the electron fluid
dominates the low-temperature
specific heat and powerfully resists the motion of 
magnetic field lines.

In this note we demonstrate that in reality the CFL phase requires equal
numbers of $u,d,s$ quarks, and
is therefore automatically
electrically neutral.   No electrons are required.
None are present, even when $\mu_e$ is nonzero.

There is precedent for such behavior.  In an ordinary superconductor,
one may consider the effect of a
perturbation
$\delta {\cal L} = \delta \mu  (e_\uparrow^\dagger  e_\uparrow 
- e_\downarrow^\dagger e_\downarrow)$ that splits 
up and down spin energies.  
It has 
long been known that
for small $\delta\mu$, the 
superconducting ground state is completely unchanged by such a
perturbation and, in particular, it contains
equal numbers of up and down spins\cite{Clogston}.   
The analogous phenomenon was recently
analyzed in QCD, in the context of
pairing between two flavors of quarks with
chemical potentials 
\begin{equation}\label{mu1mu2}
\mu_1=\bar\mu-\delta\mu\,;\ \ \mu_2=\bar\mu+\delta\mu\ .
\end{equation}
For $0<\delta\mu<\Delta_0/\sqrt{2}$, with $\Delta_0$ the superconducting
gap, the ground state of the system is precisely that obtained
for $\delta\mu=0$\cite{Bedaque,BowersLOFF}.  
By introducing $m_s\neq 0$ in the Lagrangian, 
we are
considering a rather different type of perturbation, varying the relative
mass of the paired components,
which changes (for example) the relative
velocities within a pair.  
This perturbation does change the form of the
superconducting ground state; nevertheless, the
number of particles of different types remains equal.

The argument relies on the fact that 
the CFL
ordering 
involves pairing between quarks with equal and opposite
momenta and with
different flavors. 
For such pairing to be
maximally effective, the Fermi surfaces
for different flavors of quarks must occur at equal magnitudes of the 
{\it momentum\/} (as opposed to
energy).   To put it vividly, the Fermi surfaces are rigidly locked in
momentum space.  
Since the number of occupied states, given by the occupied
volume in momentum space, is now the same for all quark
flavors, so is their number density.
Deviation 
from $N_u=N_d=N_s$ would reduce the free energy
in the absence of interactions; CFL pairing, however, reduces the 
free energy most strongly if $N_u=N_d=N_s$, and this equality
is therefore enforced.

We now make the argument concrete. We work in 
a model
in which quarks interact via a four-fermion interaction
which
we
take to be that with the quantum numbers of single-gluon
exchange\cite{ioffe}.  The argument is sufficiently general,
however, that it applies qualitatively (and quantitatively
if $\Delta_0/\bar\mu$ is small) 
to any model with four-fermion interactions which are attractive
in the appropriate channel and 
to QCD at asymptotically high density, where quarks interact
by single-gluon exchange.

We can demonstrate the physics of interest by focussing
on pairing between, say, red up quarks and green strange quarks.
Let us call these two species of quarks ``1'' and ``2'', assume
they have masses $m_1=0$ and $m_2=m_s$, and denote their 
chemical potentials as in (\ref{mu1mu2}).  
The generalization of the derivation of the gap equation for $\Delta_0$ 
given in Section 4.3 of Ref. \cite{ioffe} to the
case with $m_s\neq 0$ and $\delta\mu\neq 0$ is straightforward, although
the algebra is somewhat involved. The result is
\begin{equation}\label{BCSgap}
\frac{\Delta_0}{4G}=\int \frac{d^4 p}{(2\pi)^4}
\frac{\Delta_0 w}
{w^2-
(4\bar\mu^2 - m_s^2)p^2 - (\bar\mu+ip_0-\delta\mu)^2m_s^2},
\end{equation}
where $w\equiv \Delta_0^2+\bar\mu^2+p^2+(p_0+i\delta\mu)^2$, where 
$p\equiv |\vec p|$, where the four-fermion coupling constant $G$
is normalized as in Ref. \cite{ioffe}, and where we have
chosen to work in Euclidean space.
This gap equation 
has been derived and solved explicitly with $\delta\mu=0$
and $m_s\neq 0$ in Refs. \cite{ABR2+1,gapless} and with $m_s=0$
and $\delta\mu\neq 0$ in Ref. \cite{BowersLOFF}.
All we will need, however, is the fact that
upon replacing the integration variable $p_0$ by $p_0'=p_0+i\delta\mu$,
the gap equation is seen to be independent of $\delta\mu$. This argument
holds as long as the shift from $p_0$ to $p_0'$ does not move a pole from the
upper-half plane to the lower-half plane, or vice versa.

We wish to obtain the number densities by differentiating
the free energy density $\Omega$ with respect to $\mu$,
and so must construct $\Omega$.
For noninteracting quarks with Fermi momenta $\nu_1$
and $\nu_2$, 
\begin{eqnarray}\label{Omegafree}
\Omega_{\rm free}\left(\mu_1,\mu_2,\nu_1,\nu_2\right) &=&
\frac{1}{\pi^2}\int_0^{\nu_1}p^2(p-\mu_1)dp\nonumber\\
& &\hspace{-0.5in}+
\frac{1}{\pi^2}\int_0^{\nu_2}p^2\left(\sqrt{p^2+m_s^2}-\mu_2\right)dp.
\end{eqnarray}
We assert that the free energy of the BCS state is
\begin{eqnarray}\label{OmegaBCS}
\Omega_{\rm BCS}&=& \Omega_{\rm free}\left(\mu_1,\mu_2,\nu_1,\nu_2\right)
\nonumber\\ 
&+& \int_0^{\Delta_0} d\Delta 
\left(-\frac{2 \Delta}{G} + 8\int \frac{d^4p}{(2\pi)^4} 
{\rm ~integrand}\right)
\end{eqnarray}
where the ``integrand'' is that on the 
right hand side of the gap equation (\ref{BCSgap}), 
where $\Delta_0$ solves (\ref{BCSgap}), and where 
\begin{equation}\label{nuchoice}
\nu_1=\nu_2=\bar\mu-\frac{m_s^2}{4\bar\mu}.
\end{equation}
That is, we assert that the correct way to construct the
BCS state is to first fill noninteracting quark states
up to the common Fermi momentum (\ref{nuchoice}), and to then pair.
(The last term in (\ref{OmegaBCS}) is the condensation
energy\cite{ioffe}.)
Why is the procedure embodied in (\ref{OmegaBCS}) 
correct?  Thinking of $\nu_1$ and $\nu_2$ as
variational parameters, under what circumstances does
(\ref{nuchoice}) minimize  
the free energy (\ref{OmegaBCS})? 
Note first that if  
$\nu_1=\nu_2=\nu$ is imposed, 
then the choice (\ref{nuchoice}) for $\nu$ minimizes $\Omega_{\rm free}$, 
and therefore minimizes $\Omega_{\rm BCS}$ because the
condensation energy is independent of $\nu$.
Now, what about variation with respect to $\delta\nu\equiv\nu_2-\nu_1$?
Trying $\delta\nu>0$ reduces $\Omega_{\rm free}$, but 
exacts a cost in reduced
condensation energy:  
if $\nu_2>\nu_1$ there is a region of momentum space
$\nu_1<p<\nu_2$ wherein pairing is impossible. 
(The pair
creation operator tries to create either one quark of each
type or one hole of each type, and there
is already a ``1''-quark and a ``2''-hole at every point
in this region of momentum space. 
See
Ref. \cite{BowersLOFF} for further analysis of such ``blocking
regions''.) 
To lowest order in $m_s^2/\bar\mu^2$, $\delta\mu/\bar\mu$ 
and $\Delta_0/\bar\mu$, the free energy cost of increasing 
$\delta\nu$ from zero
outweighs the free energy benefit if 
\begin{equation}\label{localminimum}
\left|\frac{m_s^2}{4\bar\mu}-\delta\mu\right| < \Delta_0\ .
\end{equation}
This condition 
can alternatively be derived via
analysis of the locations of the poles in the 
integrand in (\ref{BCSgap}): the requirement (\ref{localminimum}) {\it is} 
the requirement that 
shifting
$p_0\rightarrow p_0'$ in (\ref{BCSgap})
not move a pole across the real axis.

The condition (\ref{localminimum})  has 
a simple interpretation: when the free energy
gained either by converting a strange quark near the common Fermi surface
into a light quark ($m_s^2/2\bar\mu-2\delta\mu$) or
by converting a light quark into a strange
quark ($2\delta\mu-m_s^2/2\bar\mu$)
compensates for the free energy lost by
breaking a single pair ($2\Delta_0$), the paired state is unstable.
Eq. (\ref{localminimum}) is the criterion for the existence of the BCS phase 
as a {\it local} 
minimum of the free energy. To check that it 
is the global minimum, we must compare
$\Omega_{\rm BCS}$ to that for the unpaired state:
\begin{equation}
\Omega_{\rm normal}=\Omega_{\rm free}\left(\mu_1,\mu_2,\mu_1,
\sqrt{\mu_2^2-m_s^2}\right).
\end{equation}
We use the gap equation to eliminate $G$ in (\ref{OmegaBCS}),
work to lowest order in $m_s^2/\bar\mu^2$, 
$\delta\mu/\bar\mu$ and $\Delta_0/\bar\mu$,
and find 
\begin{equation}
\Omega_{\rm BCS}-\Omega_{\rm normal}=\frac{\bar\mu^2}{\pi^2}\left[\left(
\frac{m_s^2}{4\bar\mu}-\delta\mu\right)^2-\frac{\Delta_0^2}{2}\right],
\end{equation}
meaning that the BCS state is the global minimum of the free energy
if
\begin{equation}\label{globalminimum}
\left|\frac{m_s^2}{4\bar\mu}-\delta\mu\right| < \frac{\Delta_0}{\sqrt{2}}.
\end{equation}
Wherever (\ref{globalminimum}) is an equality, there is a first
order phase transition from the  BCS state to the unpaired state.
At this transition,
the Fermi surfaces 
relax to the (separated) 
values favored
in the absence of interaction.
For $m_s=0$, this agrees with previous results\cite{BowersLOFF}.

Wherever (\ref{globalminimum}) is satisfied, and the paired
state is favored, we now use the fact that the gap equation
is independent of $\delta\mu$ to conclude that $\Omega_{\rm BCS}$
of (\ref{OmegaBCS}) is independent of $\delta\mu$ and therefore
\begin{equation}
N_1=\frac{\partial \Omega_{\rm BCS}}{\partial\mu_1}=
\frac{\partial \Omega_{\rm BCS}}{\partial\mu_2} = N_2.
\end{equation}
To lowest order in $\Delta_0/\bar\mu$, 
\begin{equation}
N_1=N_2=
\frac{(\bar\mu-m_s^2/4\bar\mu)^3}{3\pi^2}+\frac{1}{4\pi^2}\frac{\partial}
{\partial\bar\mu}\bar\mu^2\Delta_0(\bar\mu)^2.
\end{equation}
Thus, $N_1=N_2$ in the paired state even when $m_s\neq 0$ and 
$\delta\mu\neq 0$!
Note that $\Delta_0$, $N_1$ and $N_2$ all depend on $m_s$. The point
is that $N_1=N_2$, independent of $m_s$.
The dependence of $\Delta_0$ on $m_s$ has been analyzed
previously\cite{ABR2+1,SW2}; the only reason that
these authors failed to notice that $N_1=N_2$
is that they did not calculate $N_1$ and $N_2$.

The complete analysis of the CFL state with $m_s\neq 0$
requires the $9\times 9$ block-diagonal
color-flavor matrices given in Ref. \cite{ABR2+1}.
The analysis is more involved, but the conclusion
generalizes:  the number densities of all nine quarks
(three colors and three flavors) are the same in the CFL
phase, even when $m_s$ and $\delta\mu=\mu_e/2$
are both nonzero.  The excitations in the CFL phase
include charged Nambu-Goldstone bosons \cite{CFLmesons}, but
this does not change the analysis of electrically neutral
bulk matter: adding equal numbers of electrons and 
positively charged mesons costs free energy and is not favored.
We conclude that quark
matter in the CFL phase is electrically neutral in
the absence of any electrons.  
Even an imposed $\mu_e$
cannot push electrons into the quark matter, because introducing
electrons while maintaining charge neutrality and weak equilibrium
costs too much pairing energy.

As an example, take $m_s=200$~MeV and
consider quark matter with
$\bar\mu=400$~MeV and $\mu_e=2\delta\mu=150$~MeV.
(Contact with ordinary nuclear matter would impose $\mu_e>0$.) 
According to (\ref{globalminimum}), 
this quark matter, even with such a
large $\mu_e$, 
is in the electron-free CFL phase 
as long as $(\Delta_0/\sqrt{2})>|25-75|=50$~MeV. 
The value of $\Delta_0$ is uncertain, but 
$\Delta_0\sim 100$~MeV is not unreasonable\cite{ioffe}.
Because the stresses imposed on the CFL phase by $m_s\neq 0$
and by $\mu_e>0$ 
have opposite sign, it is more
likely than previously thought
that, if present, quark matter within neutron stars is 
in the CFL phase.

The criterion which defines
the region wherein the CFL phase is favored will
deviate somewhat from (\ref{globalminimum}).
First,
there is now an electronic contribution to 
$\Omega_{\rm normal}$ (but not to $\Omega_{\rm BCS}$!).
This contribution is only of order 
$\mu_e^4$. Second, instead of
comparing the CFL phase to a phase with no pairing,
we should compare it to phases with less symmetric 
pairing. The CFL vs. 2SC comparison of  
Ref. \cite{ABR2+1} does this at $\mu_e=0$, 
and the value of $m_s$ at which the unlocking transition 
occurs is
in good agreement with (\ref{globalminimum}),
demonstrating that,  at least at $\mu_e=0$,
this is not a large effect.
Third, we expect that, as has been demonstrated at $m_s=0$\cite{BowersLOFF},
there is a region of $(m_s,\mu_e)$ 
just outside the CFL region (\ref{globalminimum}) 
where crystalline color superconductivity,
the analogue of the Larkin-Ovchinnikov-Fulde-Ferrell (LOFF) state
in ordinary superconductivity\cite{LOFF}, is favored.
Outside the CFL region, very weak 
pairing among like flavor quarks is also possible\cite{Schaefer1Flav}.  
Because both LOFF and single-flavor condensates have much
less condensation energy than the CFL phase;
their effect on the location of the 
boundary (\ref{globalminimum}) is negligible.

We have demonstrated that
while CFL ordering is maintained, 
there will be strictly equal numbers
of the three types of quarks and
rigorous electrical neutrality, in the  
absence of any electrons.  If $m_s$ or $\mu_e$ becomes
too large, however, some less symmetric phase of quark
matter will have lower free energy.  
A first order phase transition occurs
at this boundary as the rigidly locked Fermi surfaces spring
free under accumulated tension. 

The enforced neutrality of the color-flavor locked phase
has many consequences:

In the CFL phase, there is an unbroken $U(1)_{\tilde Q}$ gauge symmetry
and a corresponding massless
photon given by a specific linear combination of the 
ordinary photon and one of the gluons\cite{CFL,ABRflux}.  
With respect to the $\tilde Q$-charge 
associated with this unbroken $U(1)$ symmetry, CFL quark
matter is electrically neutral at zero temperature, and
is a perfect insulator.  Because
of the absence of $\tilde Q$-charged excitations,  CFL quark
matter is transparent to $\tilde Q$-photons and any
$\tilde Q$-magnetic flux can move
unimpeded. 
In contrast, because electrons have nonzero $\tilde Q$-charge,
if they were present 
they would scatter
$\tilde Q$-photons and would make the material
a very good conductor in which $\tilde Q$-magnetic flux 
would be frozen in place\cite{ABRflux}.  
The absence of electrons therefore changes the conclusions
of Ref. \cite{ABRflux}: 
quark matter in the 2SC phase in
the core of a neutron star anchors magnetic fields, as described
in  Ref. \cite{ABRflux}; quark matter in the CFL phase, however,
is electron-free and therefore
offers no resistance to the motion of 
$\tilde Q$-magnetic flux as the neutron star spins down.

Similarly, if electrons were present
they would dominate the specific heat, which plays
a role in the cooling of neutron stars\cite{Blaschke,Page}.
In the absence of electrons, the specific heat at zero temperature 
is that of a neutral superfluid,
much less than previously thought. 
The qualitative conclusion that the cooling of a neutron star with a
CFL quark matter core is dominated by the (large) heat capacity
of the nuclear matter mantle remains, however.

If neutron stars have CFL cores,
the absence of electrons and consequent reduction in specific
heat and increase in transparency amplifies effects
(described in Ref.~\cite{CarterReddy})
imprinted on the time distribution of the neutrinos from
a supernova
by a transition from quark-gluon plasma to CFL quark matter 
as the hot, seconds-old protoneutron star cools. 
Effects of the first order nature of the transition
need further investigation, however.

Sch\"afer realized that
if electrons were required to maintain charge neutrality,
an alternative 
would be a condensate of negatively
charged kaons in the CFL phase\cite{SchaeferKCond}.
With $N_u=N_d=N_s$ in the CFL phase,
however, we need neither kaon condensation nor
a fluid of electrons.

The broader lesson is that at temperatures
which are nonzero and small compared to $\Delta_0$, the transport
and response properties of CFL quark matter, 
in the real world with nonzero $m_s$, are dominated
by the lightest excitations of the CFL quark matter itself,
and not  by electrons as had previously been assumed.
These bosonic degrees of freedom are the massless neutral
Nambu-Goldstone boson  associated with spontaneous baryon number violation
(superfluidity) and the neutral and charged pseudo-Nambu-Goldstone
bosons associated with spontaneous chiral symmetry breaking\cite{CFL},
which have masses of order $\sqrt{m_s m_{u,d}}\Delta_0/\bar\mu$, of order
ten MeV\cite{CFLmesons,ioffe}.  The effective field theory
which describes these light degrees of freedom (and thus, we now see,
the phenomenology) is known and at high enough density
all coefficients in
it can be determined by controlled, weak-coupling
calculations\cite{CFLmesons,ioffe}. 

Finally, the transition from an ordinary nuclear matter mantle 
to a quark matter core at some radius within a neutron star
may be greatly simplified if the transition occurs directly 
to quark matter in the CFL phase.  With a noninteracting
quark matter core, one has to face the fact that
at any given $\bar\mu$, electrically neutral nuclear matter
and electrically neutral quark matter generically have
different values of $\mu_e$. Since 
$\mu_e$ must be continuous across any interface, 
a mixed phase region is thought to form, 
within which positively charged nuclear
matter and negatively charged quark matter with the same $\mu_e$ 
coexist at any given 
radius, with $\mu_e$ changing with radius\cite{Glendenning}.
If the quark matter is in the CFL phase,
there is another possibility. 
At the $\bar\mu$ (i.e. the radius) at which $\Omega_{\rm CFL}$
crosses $\Omega_{\rm nuclear}$,
an interface between bulk nuclear matter with $N_e\neq 0$ and bulk CFL
quark matter with $N_e = 0$ may be stable, as long as the $\mu_e$ 
at the interface satisfies 
(\ref{globalminimum}). The CFL insulator cannot admit electrons
while remaining neutral, even when in equilibrium
with nuclear matter with large $\mu_e$.  
The description of
the interface is more complicated than that of the bulk phases.
Boundary layers within which local electric neutrality is not
maintained and across which an electrostatic potential
gradient develops are required, as at an ordinary metal-insulator 
boundary.

We acknowledge very
helpful conversations with 
M.~Alford, J.~Bowers, J.~Kundu, S.~Reddy, T.~Sch\"afer, E.~Shuster, 
D.~Son and M.~Stephanov.
Research supported in part by the DOE under agreement DE-FC02-94ER40818.
Work of KR supported in part by a DOE OJI grant and by the 
A. P. Sloan Foundation. Preprint MIT-CTP-3055.



\begin{references}

\bibitem{FreedmanMcLerran}
B.~A.~Freedman and L.~D.~McLerran,
Phys.\ Rev.\  {\bf D16}, 1169 (1977).



\bibitem{CFL}
M. Alford, K. Rajagopal and F. Wilczek, Nucl. Phys. {\bf B537}, 443 (1999). 


\bibitem{OtherCFL}
R.~Rapp, T.~Sch\"afer, E.~V.~Shuryak and M.~Velkovsky,
Annals Phys.\  {\bf 280}, 35 (2000);
T.~Sch\"afer,
Nucl.\ Phys.\  {\bf B575}, 269 (2000);
I.~A.~Shovkovy and L.~C.~Wijewardhana,
Phys.\ Lett.\  {\bf B470}, 189 (1999);
N.~Evans, J.~Hormuzdiar, S.~D.~Hsu and M.~Schwetz,
Nucl.\ Phys.\  {\bf B581}, 391 (2000).

\bibitem{ioffe}
For a review, see
K. Rajagopal and F. Wilczek, hep-ph/0011333.


\bibitem{CFLmesons}
R.~Casalbuoni and R.~Gatto,
Phys.\ Lett.\  {\bf B464}, 111 (1999);
D.~T.~Son and M.~A.~Stephanov,
Phys.\ Rev.\  {\bf D61}, 074012 (2000);
erratum, {\it ibid.} {\bf D62}, 059902 (2000);
D.~K.~Hong, T.~Lee and D.~Min,
Phys.\ Lett.\  {\bf B477}, 137 (2000);
C.~Manuel and M.~H.~Tytgat,
Phys.\ Lett.\  {\bf B479}, 190 (2000);
M.~Rho, E.~Shuryak, A.~Wirzba and I.~Zahed,
Nucl.\ Phys.\  {\bf A676}, 273 (2000);
K.~Zarembo,
Phys.\ Rev.\  {\bf D62}, 054003 (2000);
S.~R.~Beane, P.~F.~Bedaque and M.~J.~Savage,
Phys.\ Lett.\  {\bf B483}, 131 (2000);
D.~K.~Hong, Phys. Rev. {\bf D62}, 091501 (2000);
C. Manuel and M. Tytgat, hep-ph/0010274;
R.~Casalbuoni, R.~Gatto and G.~Nardulli,
hep-ph/0010321.



\bibitem{ABR2+1}
M.~Alford, J.~Berges and K.~Rajagopal,
Nucl.\ Phys.\  {\bf B558}, 219 (1999).


\bibitem{SW2}
T.~Sch\"afer and F.~Wilczek,
Phys.\ Rev.\  {\bf D60}, 074014 (1999).


\bibitem{Clogston}
A. M. Clogston, Phys. Rev. Lett. {\bf 9}, 266 (1962); B. S. Chandrasekhar,
App. Phys. Lett. {\bf 1}, 7 (1962).

\bibitem{Bedaque}
P.~F.~Bedaque,
hep-ph/9910247.

\bibitem{BowersLOFF}
M. Alford, J. Bowers and K. Rajagopal, hep-ph/0008208.



\bibitem{gapless}
M.~Alford, J.~Berges and K.~Rajagopal,
Phys.\ Rev.\ Lett.\  {\bf 84}, 598 (2000).


\bibitem{LOFF}
A. I. Larkin and Yu. N. Ovchinnikov,  Zh. Eksp. Teor. Fiz. {\bf 47}, 
1136 (1964) [Sov. Phys. JETP {\bf 20}, 762 (1965)];
P. Fulde and R. A. Ferrell, Phys. Rev. {\bf 135}, A550 (1964).

\bibitem{Schaefer1Flav}
T. Schaefer, Phys. Rev. {\bf D62}, 094007 (2000).

\bibitem{ABRflux}
M.~Alford, J.~Berges and K.~Rajagopal,
Nucl.\ Phys.\  {\bf B571}, 269 (2000).

\bibitem{Blaschke}
D.~Blaschke, T.~Klahn and D.~N.~Voskresensky,
Astrophys. J. {\bf 533}, 406  (2000); 
D.~Blaschke, H.~Grigorian and D.~N.~Voskresensky,
astro-ph/0009120.

\bibitem{Page}
D.~Page, M.~Prakash, J.~M.~Lattimer and A.~Steiner,
Phys.\ Rev.\ Lett.\  {\bf 85}, 2048 (2000).



\bibitem{CarterReddy}
G.~W.~Carter and S.~Reddy,
Phys.\ Rev.\  {\bf D62}, 103002 (2000).

\bibitem{SchaeferKCond}
T.~Sch\"afer,
nucl-th/0007021.

\bibitem{Glendenning}
N. K. Glendenning, Phys. Rev. {\bf D46}, 1274 (1992).



\end{references}
\end{document}